# Predicting Thermal Stress and Failure Risk in Monoblock Divertors Using 2D Finite Difference Modeling and Gradient Boosting Regression for Fusion Energy Applications

Ayobami Daramola[a]

[a]Supa, School of Physics and Astronomy. Centre of Science at Extreme Conditions. The university of Edinburgh, United Kingdom.
Email address: adaramo2@ed.ac.uk

## Abstract

This study presents a combined approach using a 2D finite difference method and Gradient Boosting Regressor (GBR) to analyze thermal stress and identify potential failure points in monoblock divertors made of tungsten, copper, and CuCrZr alloy. The model simulates temperature and heat flux distributions under typical fusion reactor conditions, highlighting regions of high thermal gradients and stress accumulation. These stress concentrations, particularly at the interfaces between materials, are key areas for potential failure, such as thermal fatigue and microcracking. Using the GBR model, a predictive maintenance framework is developed to assess failure risk based on thermal stress data, allowing for early intervention. This approach provides insights into the thermomechanical behavior of divertors, contributing to the design and maintenance of more resilient fusion reactor components.

## 1. Introduction

Nuclear fusion holds promise as a sustainable and clean energy source, with the potential to meet the growing global energy demand [1]. One of the key components in achieving stable and efficient fusion reactions is the divertor, which plays a critical role in managing the intense heat and particle flux produced during fusion [2–5]. The divertor helps maintain plasma purity and protects reactor walls, ensuring the overall performance and efficiency of the reactor. Despite robust materials, divertor components face thermal fatigue, cracking, and degradation due to extreme heat loads and particle bombardment [6].

Recent advances in computational modelling and machine learning have opened new opportunities to tackle these challenges by enhancing the ability to simulate and predict the behaviour of reactor components under extreme conditions [7–11]. In particular, the concept of a digital twin has emerged as a transformative technology in the predictive modelling of plasma-facing components (PFCs) [12]. A digital twin serves as a real-time, virtual replica of the physical system, allowing engineers to continuously monitor and predict the system's behaviour based on current operational data [12, 13]. This approach enables proactive maintenance and timely interventions, potentially extending the lifespan of critical reactor components and enhancing reactor safety.

The monoblock divertor, primarily constructed from tungsten, is designed to withstand high thermal and mechanical stresses due to its high melting point and resistance to plasma erosion [3,

5, 14–19]. However, the harsh environment, characterized by heat fluxes that can reach up to 20 MW/m², imposes significant thermal gradients that induce mechanical stresses across different material layers. This often leads to fatigue, cracking, and other types of material failure, particularly in regions with differing thermal expansion coefficients [5, 6, 14, 19–21]. Understanding these complex thermal and mechanical interactions is crucial for improving the reliability and safety of fusion reactors as they transition from experimental setups to commercial applications.

This research presents a framework that combines finite difference method (FDM) with machine learning (ML) to predict stress, temperature distribution, and potential failure points in monoblock divertors in real time. A Gradient Boosting Regressor (GBR) model is used here because of its effectiveness in capturing complex material behaviors that occur under high thermal gradients, such as those in fusion reactor environments [22].

The framework leverages temperature and stress data generated from FDM simulations to build a "digital twin" of the monoblock divertor. This model provides insights into stress concentrations and potential failure points, helping to improve understanding of how materials respond to extreme conditions. The main goal is to develop a predictive model of thermomechanical behaviour in divertors, with real-time insights into stress and heat flux distributions. Such a model can support the design and maintenance of more reliable and resilient fusion reactor components.

## 2. Methods

### 2.1. Computational and algorithm approach

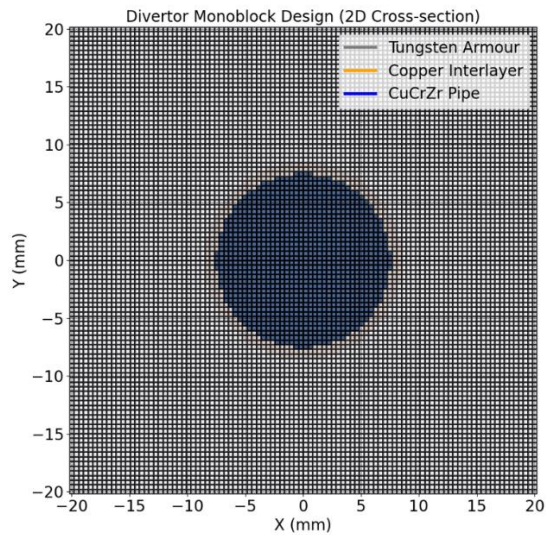

Figure 1: Cross-sectional 2D representation of the monoblock divertor model. The model includes layers of tungsten armor (grey), a copper interlayer (orange), and a CuCrZr pipe (blue). The grid illustrates the discretization for finite difference temperature and stress calculations.

In this work, a 2D cross-sectional model was chosen over a full 3D model due to the significant reduction in computational cost, allowing for faster simulations while still capturing critical planar thermal stress patterns within the divertor. The materials included are tungsten (W), copper (Cu), and a CuCrZr alloy, each with distinct thermal and mechanical properties such as thermal conductivity ($\kappa$),, thermal expansion coefficient ($\alpha$), elastic modulus ($E$), and density ($\rho$). These properties are essential in simulating heat transfer and stress accurately, aligning with the material characteristics found in literature [3, 5, 7, 14, 15, 17, 18, 21, 23]. The model used in this work discretizes the 2D cross-sectional area into a grid along the $x$ and $y$ directions, capturing the temperature and thermal stress distribution at each grid point as shown in Figure 1. The initial temperature is set to 20°C, and a 10 MW/m² heat flux is applied to simulate typical fusion reactor conditions following [3, 7].

### 2.1.1 Heat conduction calculation

Heat conduction within the divertor structure is governed by the 2D heat conduction equation, which assumes isotropic material properties and a steady-state condition:

$$\frac{\partial T(x,y,t)}{\partial t} = \tau \left( \frac{\partial^2 T}{\partial x^2} + \frac{\partial^2 T}{\partial y^2} \right) \qquad (1)$$

where $\tau = \frac{\kappa}{\rho c}$ represents thermal diffusivity. This equation captures the heat transfer over time under these assumptions, focusing on the evolution of temperature distribution across the 2D plane.

### *2.1.2 Finite difference method (FDM) for numerical solution*

The finite difference method (FDM) is employed to numerically solve the heat conduction equation. The 2D domain is divided into a uniform grid with spacings $\Delta x$ and $\Delta y$, while time is incremented in steps $\Delta t$.

(i) *Discretization*: Each grid point $(i, j)$ represents a unique position where temperature and stress are calculated.

(ii) *Temperature calculation*: The temperature at each grid point is computed iteratively at each time step as:

$$T_{ij}^{n+1} = T_{ij}^n + \Delta t \cdot \tau \left( \frac{T_{i+1,j}^n + 2T_{i,j}^n + T_{i-1,j}^n}{\Delta x^2} + \frac{T_{i,j+1}^n + 2T_{i,j}^n + T_{i,j-1}^n}{\Delta y^2} \right) \qquad (2)$$

This iterative calculation provides a detailed view of the temperature flow through the structure over time.

### 2.1.3 Thermal stress calculation

Thermal stress $\sigma_T$ is calculated from the temperature field using:

$$\sigma_T = E_T \cdot \alpha_T (T - T_0) \tag{3}$$

where $E_T$ is the elastic modulus at the temperature, $\alpha_T$ is the thermal expansion coefficient, , $T$ is the local temperature, and $T_0$ is the reference temperature set to 20°C. This approach assumes linear elastic behavior, facilitating stress calculation directly from temperature gradients.

### 2.1.4 Incorporating random fluctuations

To model the impact of environmental variability, Gaussian noise $\eta$ is introduced to the temperature values:

$$T_{ij}^{n+1} = T_{ij}^{n+1} + \eta \tag{4}$$

where $\eta \sim N(0, \sigma^2)$ with $\sigma$ calibrated from empirical reactor data. This randomization enhances the model's robustness by simulating the natural fluctuations expected in a reactor environment.

## 2.2. Gradient boosting regression (GBR) for failure prediction

To assess the risk of thermal failure, thermal stress data from the FDM simulations is analyzed using a machine learning approach. A Gradient Boosting Regressor (GBR) is chosen for its ability to capture complex non-linear relationships and incremental learning advantages over other models like Random Forest or Support Vector Machines, particularly given the structured nature of the data and need for fine-tuned failure predictions.

### 2.2.1 Database construction and model (input and output)

The dataset for training the GBR model is constructed from FDM simulation outputs. Features at each grid point and time step include:

(i) **Temperature** ($T$) at each grid point.

(ii) **Thermal stress** ($\sigma_T$), as derived from Equation (3).

A failure probability metric is also recorded, triggered when thermal stress exceeds a specified threshold. Additional spatial gradients of temperature and stress provide context on localized patterns and heat concentration within the divertor structure.

### 2.2.2. Model training and evaluation

The GBR model is trained and validated as follows:

(i) *Data splitting*: 80% of the data is used for training, and 20% for validation.

(ii) *Hyperparameter tuning*: Key GBR parameters, such as the number of estimators, learning rate, and tree depth, are optimized through grid search. The final settings include used n_estimators=100, learning_rate=0.1, max_depth=3 and random_state=42.

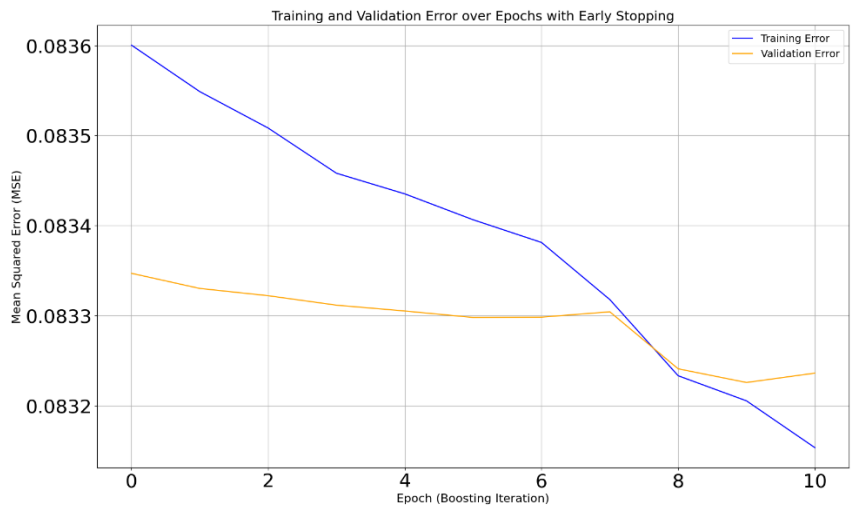

Figure 2: Training and validation error versus boosting iterations, illustrating GBR performance with early stopping applied.

(iii) *Performance metrics*: Model performance is assessed using Mean Squared Error (MSE) and R-squared values to measure accuracy and generalization capacity. Results are plotted to visualize training and validation errors over boosting iterations (Figure 2).

### 2.2.3 Model prediction and visualization

Using the trained GBR model, failure probabilities are predicted across the grid. This prediction identifies high-risk areas in the divertor structure, highlighting zones susceptible to thermal stress and potential fatigue. Visualizations display temperature distribution, thermal stress, and failure probability across the divertor layers, offering insight into which materials (W, Cu, CuCrZr) are most at risk under various thermal conditions.

# 3. Results

## 3.1 Temperature distribution

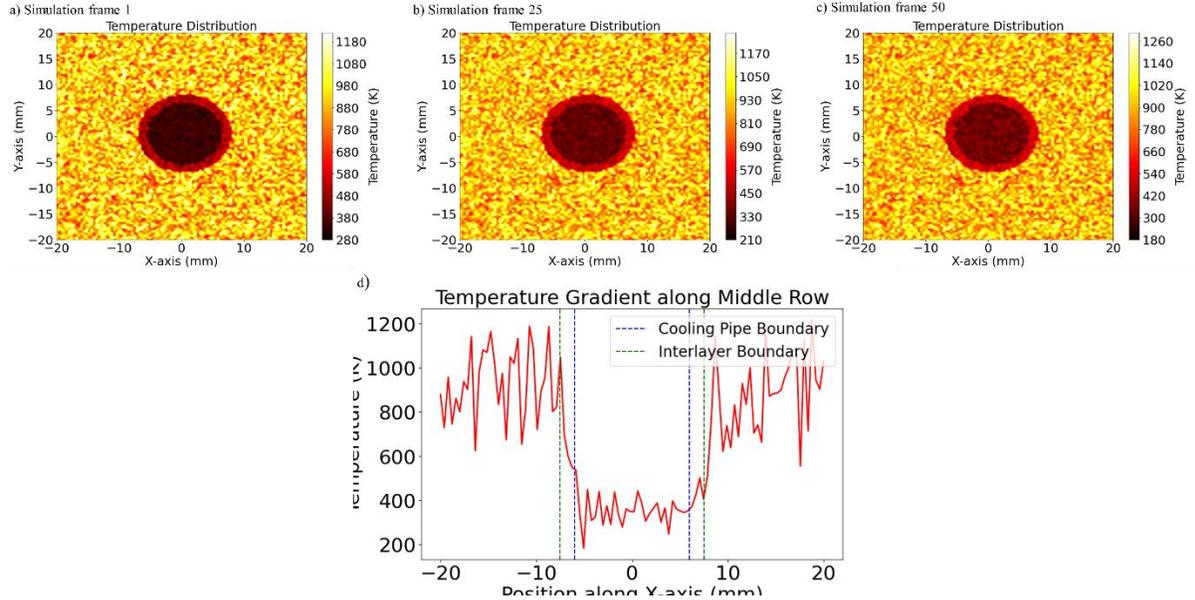

Figure 3: Temperature distribution across different simulation time frames: (a) Initial frame showing lower central temperatures, (b) 25th frame displaying partial heat dissipation, and (c) 50th frame illustrating a more uniform temperature gradient. The color scale represents temperature in Kelvin. d) Temperature gradient along the central region of the monoblock, showing sharp drops at cooling pipe and interlayer boundaries.

As shown in **Figure 3**, the temperature distribution across the divertor follows a clear pattern, with the highest temperatures occurring near the heat flux interface. The tungsten (W) armor layer, positioned at the outermost section of the model, experiences the highest temperature increase due to its exposure to the plasma-facing surface. The temperature distribution in this region reaches up to 1200°C after several iterations, as seen in **Figure 3b and 3c**, which shows the temperature at a specific time step after the heat flux is applied.

The copper interlayer (Cu) and CuCrZr pipe, being thermally conductive, exhibit a more uniform temperature profile but still experience elevated temperatures as the heat spreads inward. In Figure 3a, we observe the initial thermal gradients and how heat is conducted across the divertor structure during the early stages of simulation and later stages as shown in Figure 3d. In the region between approximately -10 mm and 10 mm (near these boundary lines), the temperature shows a sharp drop from high values (above 1000 K) to lower values (around 300-400 K). This indicates a steep gradient in temperature, especially as the temperature decreases sharply when approaching the cooling pipe boundary.

## 3.2 Heat flux distribution

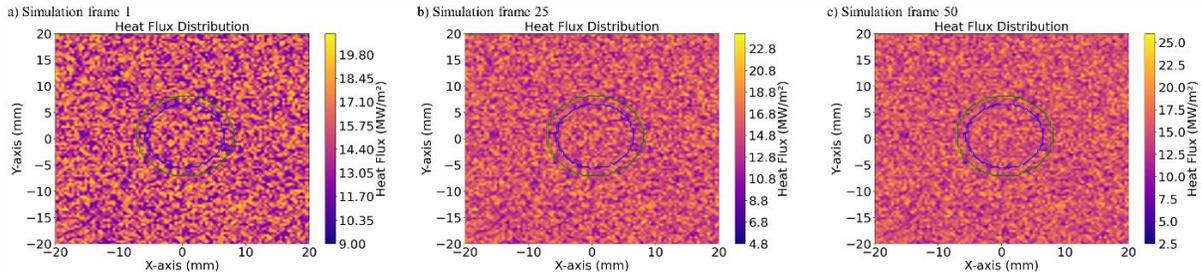

Figure 4: Heat flux distribution across selected simulation frames: (a) Initial frame showing concentrated heat flux around the central region, (b) 25th frame depicting partial diffusion of heat flux, and (c) 50th frame illustrating further spreading of heat flux and a reduction in intensity. The color scale represents heat flux in MW/m².

In the 2D divertor model, heat flux distribution across the surface varies widely, from approximately 9 MW/m² in low plasma interaction zones to peaks of 32 MW/m² in areas exposed to intense plasma, as shown in Figure 4. This distribution is influenced by the cooling pipe and interlayer design. As seen in Figures 4b and 4c, the cooling pipe effectively reduces heat flux in its immediate surroundings, helping to manage thermal loads and prevent localized overheating, consistent with the observed temperature distribution. The interlayer materials further guide heat flow: regions with higher thermal conductivity support rapid heat dissipation, while insulating layers help control localized temperatures, as observed during early stages of simulation (Figure 4a). In Figure 4a, representing initial conditions, heat concentrated near the plasma source gradually decreases as it is directed toward cooling zones. This redistribution of heat flux minimizes hotspots and reduces thermal stress on the structure during early exposure. The following section will present the corresponding thermal stress results.

## 3.3 Thermal stress accumulation

Thermal stress accumulation follows a similar trend, with stress building up most significantly in the tungsten armor. As shown in Figure 5, the thermal stress rises sharply in the first few iterations, particularly in regions where the temperature gradient is steep as shown in Figure 2d. These regions, particularly near the interface between tungsten and copper, are at risk of developing high-stress concentrations.

As the simulation progresses, stress continues to accumulate gradually. Fatigue points, where thermal stress exceeds the material's threshold for damage, begin to appear only after a certain number of iterations as shown in Figure 5b with the red spot. This delay is due to the incremental nature of thermal stress buildup, where the material gradually accumulates strain as the temperature increases. Fatigue risks emerge when the accumulated stress exceeds the material's tolerance threshold, as shown in Figure 5b-5d, which highlights these high-stress regions where failure is most likely to occur.

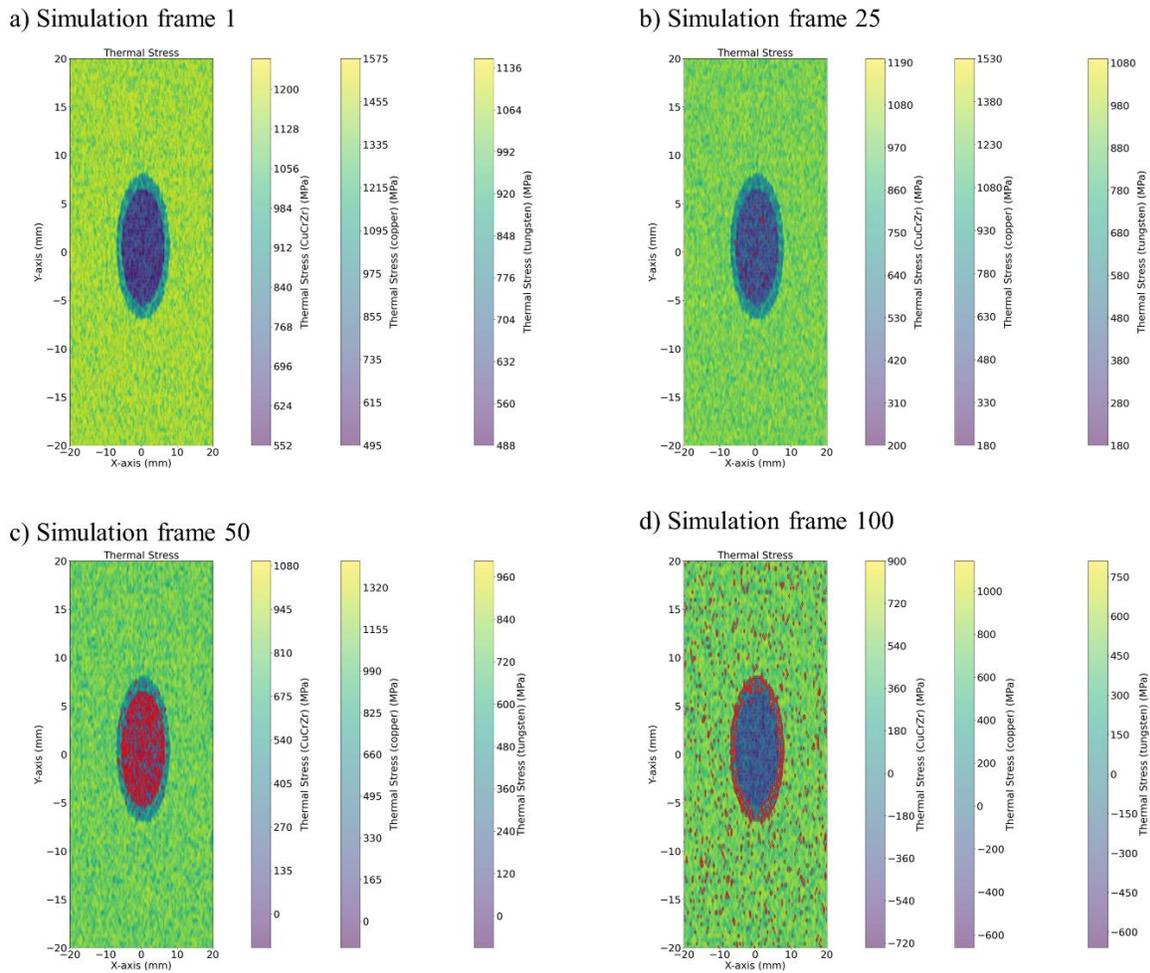

Figure 5: Thermal stress distribution over selected simulation frames: (a) Initial frame showing high thermal stress concentration at the core, (b) 25th frame illustrating the spread of thermal stress with decreasing intensity, (c) 50th frame indicating further diffusion of stress across materials, and (d) 100th frame demonstrating a notable reduction in thermal stress as the system nears equilibrium. Stress values are presented for CuCrZr, Copper, and Tungsten, with scales in MPa.

### 3.3 Stress and failure risk prediction

The GBR model, trained on the thermal stress data, successfully predicts failure probabilities across the divertor. These predictions are visualized in Figure 6, which presents three key graphical representations. As shown in Figure 6a, a bar chart displays the thermal stress levels at various grid points across the divertor structure, showing how stress levels vary by material and location. The highest bars indicate regions of potential failure risk, particularly in the Copper and CuCrZr layers.

As shown in Figure 6b, a scatter plot presents the failure probability predicted by the GBR model. Each point represents a grid point in the divertor, with the color of each point indicating the predicted risk of failure. The scatter plot highlights areas where the predicted failure probability exceeds a critical threshold, signifying regions where the divertor structure is most vulnerable to thermal damage. This information can be valuable for maintenance analysis by highlighting areas with higher risk scores. As shown in Figure 6c, the CuCrZr and copper regions show higher risk scores compared to the tungsten areas, which corresponds to the increased thermal stress levels in these materials. This correlation helps identify high-risk regions that may require early intervention.

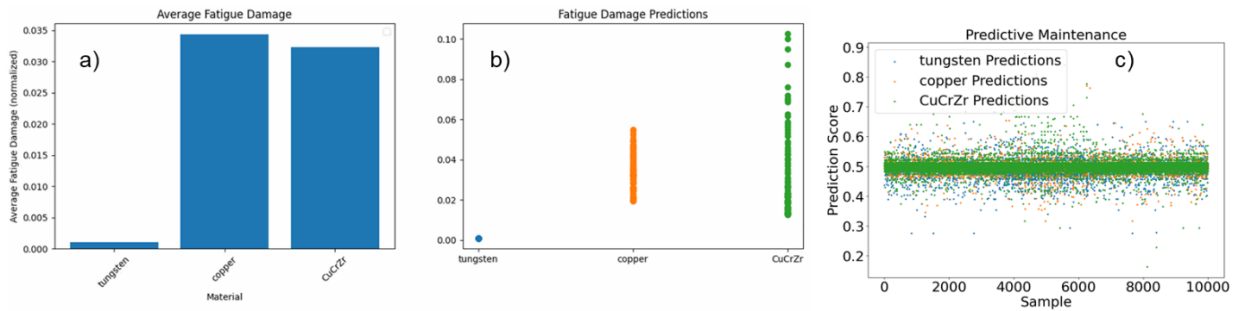

Figure 6: Fatigue damage analysis. (a) The bar chart presents average fatigue damage levels, showing higher values for CuCrZr and Copper compared to Tungsten. (b) The scatter plot illustrates individual fatigue predictions, highlighting Tungsten's relative resilience under thermal stress in comparison to CuCrZr and Copper. (c) Predictive maintenance scores for Tungsten, Copper, and CuCrZr, illustrating performance stability with scores clustering around the mean for each material.

## 4. Discussion

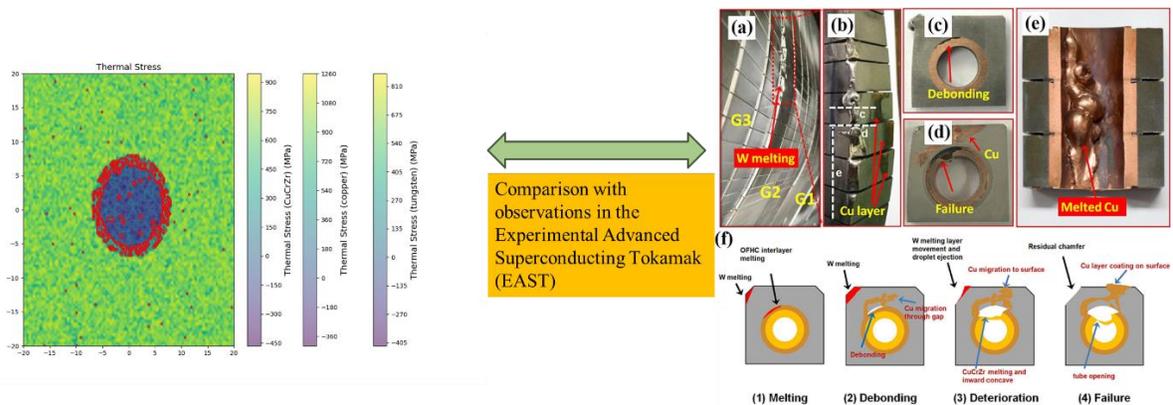

Figure 7: Comparison of simulated failure points in the 2D divertor model with experimentally observed damage in the EAST divertor under high heat flux conditions [5, 24].

The appearance of fatigue points after a certain number of iterations is a direct result of the gradual accumulation of thermal stress. Initially, the thermal stress remains below the threshold for material failure, but over time, the continuous heat flux and temperature rise cause incremental increases in stress. The fatigue points emerge when the stress builds up sufficiently to surpass the material's yield strength. This phenomenon is common in high-temperature environments where materials undergo cyclic loading, and the delayed onset of fatigue emphasizes the importance of considering long-term thermal exposure when evaluating material integrity.

The simulation results align well with experimental observations, particularly the concentration of thermal stress at material interfaces, which mirrors the degradation patterns seen in fusion reactor components under high heat flux conditions as shown in Figure 7 [5, 24].

Figures 3-6 provide clear insights into how the divertor constituent materials behaves under thermal loading. The results underscore the importance of material selection, as regions of high stress near the tungsten-copper interface are the most susceptible to failure. Additionally, the GBR model's predictions provide an early warning system for regions at risk, allowing for proactive management of the divertor's performance in future reactors.

## Conclusion

This study successfully employed a simplified 2D finite difference method in combination with machine learning to analyze thermal stress and predict failure points in monoblock divertors. The results highlight the significance of temperature gradients and material interfaces in causing thermal fatigue and cracking. The predictive maintenance framework, using Gradient Boosting Regression, provides valuable insights into high-risk areas, which could be used for early intervention and proactive maintenance. Moving forward, the model could be expanded to a 3D framework to more accurately simulate complex geometries and interactions. Additionally, exploring other machine learning models or enhancing the digital twin approach with real-time operational data could further improve the accuracy and applicability of this predictive tool in fusion energy systems.

## Data availability

The code utilized for generating the results and conducting the analysis presented in this paper is available for download at https://github.com/ayobamidaramola98/Real-time-simulation/tree/main

1. Li, X., Zhang, L., Wang, G., Long, Y., Yang, J., Qin, M., Qu, X., So, K.P.: Microstructure evolution of hot-rolled pure and doped tungsten under various rolling reductions. Journal of Nuclear Materials. 533, 152074 (2020). https://doi.org/10.1016/j.jnucmat.2020.152074
2. Zhu, D., Li, C., Gao, B., Ding, R., Wang, B., Guo, Z., Xuan, C., Yu, B., Lei, Y., Chen, J., EAST Team, the: In situ leading-edge-induced damages of melting and cracking W/Cu monoblocks as divertor target during long-term operations in EAST. Nuclear Fusion. 62, 056004 (2022). https://doi.org/10.1088/1741-4326/ac3f48
3. Domptail, F., Barrett, T.R., Fursdon, M., Lukenskas, A., You, J.-H.: The design and optimisation of a monoblock divertor target for DEMO using thermal break interlayer. Fusion Engineering and Design. 154, 111497 (2020). https://doi.org/10.1016/j.fusengdes.2020.111497
4. Nogami, S., Toyota, M., Guan, W., Hasegawa, A., Ueda, Y.: Degradation of tungsten monoblock divertor under cyclic high heat flux loading. Fusion Engineering and Design. 120, 49–60 (2017). https://doi.org/10.1016/j.fusengdes.2017.04.102

5. Xu, D., Cheng, J., Chen, P., Fu, K., Wei, B., Chen, R., Luo, L., Xu, Q.: Recent progress in research on bonding technologies of W/Cu monoblocks as the divertor for nuclear fusion reactors. Nuclear Materials and Energy. 36, 101482 (2023). https://doi.org/10.1016/j.nme.2023.101482
6. Wang, K., Xi, Y., Zan, X., Zhu, D., Luo, L., Ding, R., Wu, Y.: Failure analysis of damaged tungsten monoblock components of upper divertor outer target in EAST fusion device. Nuclear Engineering and Technology. 56, 2307–2316 (2024). https://doi.org/10.1016/j.net.2024.01.041
7. Humphrey, L.R., Dubas, A.J., Fletcher, L.C., Davis, A.: Machine learning techniques for sequential learning engineering design optimisation. Plasma Phys Control Fusion. 66, 025002 (2024). https://doi.org/10.1088/1361-6587/ad11fb
8. Mohamed, S., Po, G., Lewis, R., Nithiarasu, P.: Multiscale computational study to predict the irradiation-induced change in engineering properties of fusion reactor materials. Nuclear Materials and Energy. 39, 101647 (2024). https://doi.org/10.1016/j.nme.2024.101647
9. Piccione, A., Berkery, J.W., Sabbagh, S.A., Andreopoulos, Y.: Physics-guided machine learning approaches to predict the ideal stability properties of fusion plasmas. Nuclear Fusion. 60, 046033 (2020). https://doi.org/10.1088/1741-4326/ab7597
10. Humphreys, D., Kupresanin, A., Boyer, M.D., Canik, J., Chang, C.S., Cyr, E.C., Granetz, R., Hittinger, J., Kolemen, E., Lawrence, E., Pascucci, V., Patra, A., Schissel, D.: Advancing Fusion with Machine Learning Research Needs Workshop Report. Journal of Fusion Energy. 39, 123–155 (2020). https://doi.org/10.1007/s10894-020-00258-1
11. Pavone, A., Merlo, A., Kwak, S., Svensson, J.: Machine learning and Bayesian inference in nuclear fusion research: an overview. Plasma Phys Control Fusion. 65, 053001 (2023). https://doi.org/10.1088/1361-6587/acc60f
12. Davis, A., Waldon, C., Muldrew, S.I., Patel, B.S., Verrier, P., Barrett, T.R., Politis, G.A.: Digital: accelerating the pathway. Philosophical Transactions of the Royal Society A: Mathematical, Physical and Engineering Sciences. 382, (2024). https://doi.org/10.1098/rsta.2023.0411
13. Liu, W., Han, L., Huang, L.: Design and optimization of molten salt reactor monitoring system based on digital twin technology. Kerntechnik. 87, 651–660 (2022). https://doi.org/10.1515/kern-2022-0055
14. You, J.H., Mazzone, G., Visca, E., Greuner, H., Fursdon, M., Addab, Y., Bachmann, C., Barrett, T., Bonavolontà, U., Böswirth, B., Castrovinci, F.M., Carelli, C., Coccorese, D., Coppola, R., Crescenzi, F., Di Gironimo, G., Di Maio, P.A., Di Mambro, G., Domptail, F., Dongiovanni, D., Dose, G., Flammini, D., Forest, L., Frosi, P., Gallay, F., Ghidersa, B.E., Harrington, C., Hunger, K., Imbriani, V., Li, M., Lukenskas, A., Maffucci, A., Mantel, N., Marzullo, D., Minniti, T., Müller, A.V., Noce, S., Porfiri, M.T., Quartararo, A., Richou, M., Roccella, S., Terentyev, D., Tincani, A., Vallone, E., Ventre, S., Villari, R., Villone, F., Vorpahl, C., Zhang, K.: Divertor of the European DEMO: Engineering and technologies for power exhaust. Fusion Engineering and Design. 175, 113010 (2022). https://doi.org/10.1016/j.fusengdes.2022.113010
15. Li, M., You, J.-H.: Structural impact of creep in tungsten monoblock divertor target at 20 MW/m 2. Nuclear Materials and Energy. 14, 1–7 (2018). https://doi.org/10.1016/j.nme.2017.12.001
16. Fursdon, M., You, J.-H., Barrett, T., Li, M.: A hybrid analysis procedure enabling elastic design rule assessment of monoblock-type divertor components. Fusion Engineering and Design. 135, 154–164 (2018). https://doi.org/10.1016/j.fusengdes.2018.07.014
17. CARDELLA, A., DI PIETRO, E., BROSSA, M., GUERRESCHI, U., REALE, M., REHEIS, N., VIEIDER, G.: DESIGN MANUFACTURING AND TESTING OF THE MONOBLOCK DIVERTOR. In: Fusion Technology 1992. pp. 211–215. Elsevier (1993)
18. Richou, M., Li-Puma, A., Visca, E.: Design of a water cooled monoblock divertor for DEMO using Eurofer as structural material. Fusion Engineering and Design. 89, 975–980 (2014). https://doi.org/10.1016/j.fusengdes.2014.04.019
19. Hirai, T., Escourbiac, F., Barabash, V., Durocher, A., Fedosov, A., Ferrand, L., Jokinen, T., Komarov, V., Merola, M., Carpentier-Chouchana, S., Arkhipov, N., Kuznetcov, V., Volodin, A., Suzuki, S., Ezato, K., Seki, Y., Riccardi, B., Bednarek, M., Gavila, P.: Status of technology R&D for the ITER tungsten divertor monoblock. Journal of Nuclear Materials. 463, 1248–1251 (2015). https://doi.org/10.1016/j.jnucmat.2014.12.027

20. Li, M., You, J.-H.: Structural impact of armor monoblock dimensions on the failure behavior of ITER-type divertor target components: Size matters. Fusion Engineering and Design. 113, 162–170 (2016). https://doi.org/10.1016/j.fusengdes.2016.10.014
21. Gao, B., Ding, R., Zhang, L., Li, C., Xie, H., Zeng, L., Yang, J., Wang, B., Zhu, D., Chen, J.: Impact of W monoblock damage on EAST operations in recent campaigns. Fusion Engineering and Design. 169, 112623 (2021). https://doi.org/10.1016/j.fusengdes.2021.112623
22. Keprate, A., Ratnayake, R.M.C.: Using gradient boosting regressor to predict stress intensity factor of a crack propagating in small bore piping. In: 2017 IEEE International Conference on Industrial Engineering and Engineering Management (IEEM). pp. 1331–1336. IEEE (2017)
23. Nogami, S., Guan, W.H., Hattori, T., James, K., Hasegawa, A.: Improved structural strength and lifetime of monoblock divertor targets by using doped tungsten alloys under cyclic high heat flux loading. Phys Scr. T170, 014011 (2017). https://doi.org/10.1088/1402-4896/aa864d
24. Li, C., Zhu, D., Ding, R., Wang, B., Chen, J., Gao, B., Lei, Y.: Characterization on the melting failure of CuCrZr cooling tube of W/Cu monoblocks during plasma operations in EAST. Nuclear Materials and Energy. 25, 100847 (2020). https://doi.org/10.1016/j.nme.2020.100847